\documentclass[twocolumn,showpacs,amsmath,amssymb]{revtex4}

\usepackage{graphicx}
\usepackage{color}
\begin{document}
\title{Origin of branch points in the spectrum of \emph{PT}-symmetric periodic potentials}
\author{Ching-Hao Chang, Shi-Ming Wang, and Tzay-Ming Hong}
\affiliation{Department of Physics, National Tsing Hua University, Hsinchu 30043, Taiwan, Republic of China}
\date{\today}

\begin{abstract}
There exists multiple branch points in the energy spectrum for some \emph{PT}-symmetric periodic potentials, where the real eigenvalues turn into complex ones.
By studying the transmission amplitude for a localized complex potential, we elucidate the physical origin of the breakdown of perturbation method and Born approximation.
Mostly importantly, we derive an analytic criteria to determine  why, when and where the bifurcation will occur.

\end{abstract}
\maketitle

\section{INTRODUCTION}
Non-Hermitian (NH) quantum mechanics (QM) has been used to describe
an open system because complex eigenenergies imply the existence of
a sink or source  for the particle. The \emph{PT}-symmetric
hamiltonian is a special case in NHQM that can exhibit entirely real
spectra\cite{pt1}. Its development has gone through various
stages\cite{pt2} with applications ranging from quantum
cosmology\cite{16}, quantum field theory\cite{19} to quantum
computation\cite{brach1,brach2}. 
A flourishing example is in connection with the
optical lattice which can be simulated by the Schr\"{o}dinger
equation. This provides a fertile ground to test the NH-related
concepts experimentally. In this spirit scientists have studied the
\emph{PT}-symmetric periodic potentials and investigated several
interesting features: (1) The existence of several branch points
which demarcate the real and complex spectra\cite{dorey}. (2)
Optical soliton solutions are found within the energy gap with real
eigenvalues\cite{soliton}. (3) A quantum phase transition is
identified when a sufficiently large real periodic potential is
superimposed to render all the spectrum real\cite{beam}. The latter two, which
provide the basis for the fruitful beam dynamics\cite{soliton,
beam}, are closely related to the branch point and bifurcation
feature. Although the branch points and a complex spectrum have been
attributed to the failure of perturbation expansion\cite{visual}, we
believe it is worthwhile to study in more detail the physical origin of this failure.

We shall start by studying a  localized imaginary potential
before embarking on the more interesting periodic case. Several new
features are discussed in Section II which include (1) Existence of
an upper bound for the imaginary part of eigenenergies.
(2) Breakdown of the Born approximation for scattering states unless
a large enough real potential is added. (3) Fano resonance within the imaginary barrier
can cause the transmission amplitude to become huge
by repetitive enhancements. (4) An incoming wave packet will also
exhibit special features when its momentum is near the
resonance. We find that the size of the transmission amplitude is the key to features (2) and (3).
In Section III we argue that the spectrum of the \emph{PT}-symmetric periodic potential is predominantly shaped by the transmission amplitude for a single unit cell. Therefore, the conclusions in Section II for a localized potential are still applicable in the periodic case. By use of the Bloch's theorem, we derive an analytic criteria for the occurrence of branch point and the band structure bifurcation.
 Discussions and conclusions are
arranged in the final Section IV.

\section{SINGLE IMAGINARY SQUARE  POTENTIAL}
\subsection{Localized states}
At first glance, one might think that the eigenenergies of a complex
potential are most likely also complex. This is in fact not true.
For instance, take an imaginary square potential $iV_{0}$ from $x=0$
to $x=a$ where $V_0>0$ and zero potential elsewhere. It exhibits
both real/imaginary eigenvalues, which are continuous/discrete and
correspond to scattering/localized states, respectively.

Since the barrier potential is symmetric with respect to $x=a/2$, the spatial part of the localized wavefunction, $\psi
(x,t)=\phi(x)e^{-iEt}$ where $\hbar$ has been set to unity for
convenience, can be expressed as:
\begin{equation}
\phi(x) = \left\{\begin{array}{lll}
\pm e^{-(ik-q)(x-a/2)}, &x \le 0, \\
A \big(e^{ip(x-a/2)}\pm e^{-ip(x-a/2)}\big), & 0 \le x\le a , \\
e^{(ik-q)(x-a/2)}, &x \ge a
\end{array} \right.
\label{eq:fstate}
\end{equation}
where momentum $k$ and $q$ are real and positive, $p$ is complex, and the
$\pm$ sign denotes respectively the even/odd-parity state with respect to $x=a/2$. The minus sign in the exponent of $\phi (x\le 0)$ is to prevent it from blowing up at $x\rightarrow -\infty$. These
momenta are related by the complex energy $E$:
\begin{equation}
E=\frac{-(ik-q)^{2}}{2m}=\frac{p^2}{2m}+iV_{0}.
\end{equation}

One should be aware that the standard concept of density current need not survive\cite{pt-sq3} the transition to NHQM. For instance, in the case of a \emph{PT}-symmetric potential $j(x,t)$ has to be redefined as $ -i\big[\psi^{*}(x,t)\partial \psi (-x,t)/\partial x-\psi (-x,t)\partial \psi^{*}(x,t)/\partial x\big]/2 m$ in order to obtain the continuity equation. For the sake of arguments, let us stick to the standard notion and deduce the following mathematical equation from the Schr\"{o}dinger equation for a general complex potential $V(x)$:
\begin{equation}
2{\rm Im}V(x)|\psi (x,t)|^{2}=\frac{\partial}{\partial t}\big| \psi
(x,t)\big|^{2}+\frac{\partial}{\partial x}j_{0}(x,t)
\label{eq:continuity}
\end{equation}
where the symbol Im denotes taking the imaginary part and $j_{0}(x,t)\equiv -i\big[\psi^{*}(x,t)\partial \psi (x,t)/\partial x-\psi (x,t)\partial \psi^{*}(x,t)/\partial x\big]/2 m$. 
Integrating over the whole space gives
\begin{equation}
\int_{-\infty}^{\infty} dx\frac{\partial}{\partial t}|\psi
(x,t)|^{2}=2\int_{-\infty}^{\infty}dx{\rm Im}V(x)|\psi (x,t)|^{2}
\end{equation}
where $j_{0}(x,t)$ vanishes at $x=\pm \infty$ for
the localized state in Eq.(\ref{eq:fstate}). This leads to the following inequality
\begin{equation}
{\rm Im}E\int_{-\infty}^{\infty} dx|\psi (x,t)|^{2}<{\rm Max}({\rm
Im}V)\int_{-\infty}^{\infty} dx|\psi (x,t)|^{2} \label{eq:prof}
\end{equation}
which tells us that the height of the imaginary barrier $V_{0}$
serves as an upper bound for  ${\rm Im}E$:
\begin{align}
0\le{\rm Im}E<V_{0}. \label{eq:limit-E}
\end{align}

By matching the boundary conditions, we can further show that the
number of these localized states is finite and their energies are
discrete. 
In contrast to being zero for a real well, the current density $j(x,t)$ is
not only nonvanishing, but also quantized for an imaginary
barrier/well:
\begin{equation}
j_n(x,t)={k_n\over m}e^{-2q_n |x-{a/2}|+2{\rm Im}E_n t}
\end{equation}
where $n$ labels these discrete localized states.

All the derivations so far are based on the assumption that $V_0>0$.
However, it is easy to generalize to an imaginary well.  The action
of changing the potential from $iV_0$ to $-iV_0$ is equivalent to
reversing the time, which transforms the hydrant-like bound state
into a sink. The same conclusion applies to the eigenfunctions and
eigenenergies after we take the complex conjugate of the
time-independent Schr\"{o}dinger equation. As a result, the upper
bound in Eq.(\ref{eq:limit-E}) needs to be modified as
\begin{align}
-V_{0}<{\rm Im}E\le 0. \label{eq:2limit-E}
\end{align}

\subsection{Scattering states}
\begin{figure}[h!]
\hspace{0mm}\includegraphics[width=0.45\textwidth]{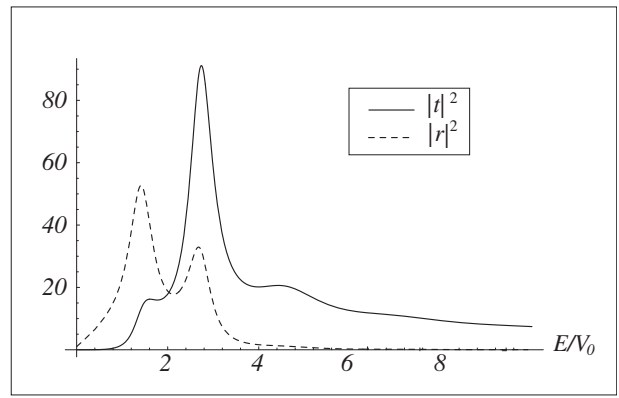}
\caption{Set the effective mass of electron as $m=0.5$.  The
transmission and reflection coefficients of an imaginary barrier
with height $V_{0}=5$ and width $a=2$ are plotted as a function of
$E/V_0$ where $E$ represents the energy of the scattering state. }
\label{f:tr0}
\end{figure}

\begin{figure}[h!]
\hspace{0mm}\includegraphics[width=0.45\textwidth]{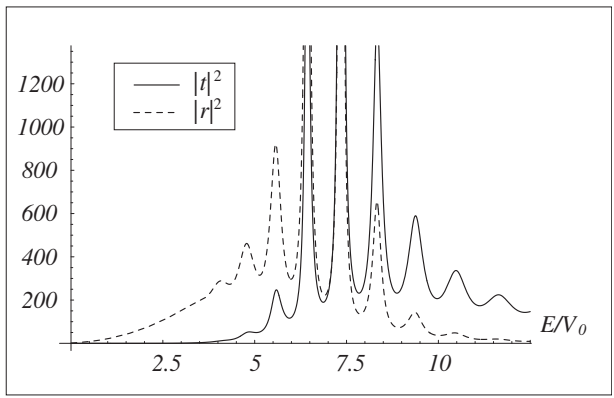}
\caption{Same as Fig.\ref{f:tr0} except the barrier height is
increased by eight folds to $V_{0}=40$.} \label{f:tr1}
\end{figure}

Another interesting problem is to study the transmission and
reflection coefficients for a complex square barrier. This is relevant to our discussion of the spectrum for a periodic potential in Section III. The scattering
state takes the form:
\begin{align}
\phi_{r,k}(x) = \left\{\begin{array}{lll}
e^{ikx} +r e^{-ikx},&x \le 0, \\
Ae^{i(p-iq)x}+Be^{-i(p-iq)x}, &0 \le x \le a , \\
te^{ikx}, &x \ge a
\end{array} \right.
\label{phi}
\end{align}
where subscript $r$ is to distinguish this right-moving state from
its degenerate state $\phi_{l,k}(x)$ that travels in the opposite
direction. Fitting the boundary conditions, we find
\begin{align}
&t=\frac{-4kk'e^{-ika}}{e^{i(k+k')a}(k-k')^{2}-e^{i(k-k')a}(k+k')^{2}}\\
&r=\frac{2i(k^{2}-k'^{2})\sin(k'a)e^{-ika}}{e^{i(k+k')a}(k-k')^{2}-e^{i(k-k')a}(k+k')^{2}}
\end{align}
where $k'\equiv p-iq$ is the complex momentum inside the barrier.
The amplitude $t$ in Eq.(\ref{phi}) consists of contributions from
all paths that reflect inside the barrier arbitrary number of times:
\begin{equation}
t=t_{1} t_{2} e^{ik'a}+t_{1} r_{2} r_{1} t_{2} e^{3ik'a}+t_{1} r_{2}
r_{1} r_{2} r_{1} t_{2} e^{5ik'a}+.... \label{t}
\end{equation}
where $t_{1,2}$/$r_{1,2}$ denote the transmission/reflection
amplitudes at the boundaries $x=0$ and $x=a$, respectively. The
higher-order terms in Eq.(\ref{t}) converge to zero for a real
barrier. But this is no longer true for a positive imaginary barrier
because the longer they stay in the source the more probability they
can gain. As a result, the values of resonance peaks can be much
bigger than unity, and the Born approximation that neglects the
higher-order terms is expected to break down. These two features are
clearly shown in Fig.\ref{f:tr0} and the number of resonances and
their peak values increase when we raise $V_0$ in Fig.\ref{f:tr1}.

Another important issue of the scattering process is the
interference condition for Fano resonance. In real barriers, the
resonance in transmission results from the constructive interference
between different scattering paths
 shown in Eq.(\ref{t}), which gives the familiar condition
$a=n\lambda/2\equiv n\pi/p$. In the mean time, reflection peaks are
attributed to the destructive interference when
\begin{equation}
a=(n+{1\over2}){\lambda\over 2} \label{n}
\end{equation}
is satisfied. These two conditions become more complicated for an
imaginary potential because the phase shift due to reflections from
the boundaries between the imaginary and real potential regimes is
energy-dependent. The amplitude $r_{1,2}$ in Eq.(\ref{t}) can be
determined by studying the reflection by an imaginary step potential
of height $iV_0$:
\begin{align}
r_{1,2} =\frac{k-k'}{k+k'}\simeq \left\{\begin{array}{ll}
1, &E \ll V_{0} \\
i{V_{0}}/{4E},   &E \gg V_{0}
\end{array} \right.
\label{phase}
\end{align}
which shows that there is an additional phase shift ranging from 0 to $\pi/2$ as $E$ increases. It is
straightforward to write down an equation similar to Eq.(\ref{t}) for
the reflection amplitude, $r$. Since both equations are geometric
series with the same common ratio, $r_1 r_2 \exp (2ik'a)$, it is
expected that the transmission and reflective resonances for an
imaginary barrier should all follow the same second criterion,
Eq.(\ref{n}), at the limit $E \gg V_{0}$. To verify these interference
features, a partial list of the energies, peak values of the transmission coefficient and
corresponding $n$ in Fig.\ref{f:tr1} are numerically calculated and
summarized in Tab.~\ref{tb:table}. For an imaginary well,
Eq.(\ref{phase}) remains valid after $V_0$ is replaced by $-V_0$.
However, the resonance peaks are no longer pronounced because the
sink now draws in probability as the wave reflects in the well.
\begin{table}[tbp]
\begin{tabular}[t]{|c|c|c|c|c|}
\hline
$E$ & 223.6 & 257.4 & 294.1 & 333.4\\
\hline
$|t|^{2}$ & 247.5 & 2322.5 & 24485.9 & 1507.7\\
\hline
$n$ & 13.52 & 14.49 & 15.48 & 16.47\\
\hline
\end{tabular}
\caption{Partial list of the energies and peak values of the
transmission coefficient in Fig.\ref{f:tr1} and their corresponding
$n$ defined by $n=2a/\lambda$.} \label{tb:table}
\end{table}

In order to facilitate the discussion of a \emph{PT}-symmetric
periodic potential in Section III, let us put the square barrier and
well side by side to form a simple \emph{PT}-symmetric potential.
The perturbation expansion can be shown to still break down as in the periodic case\cite{beam}. The similarity goes further that the perturbation in both potentials can be
salvaged by incorporating a big-enough real potential. On the other hand, we can ask how strong a complex potential can ruin the perturbation expansion. Take this pair of square barrier and well potential,  $V(x)=\{U_{0}[\Theta(x+a)-\Theta(x-a)]+iV_{0}[\Theta(x+a)-\Theta(x)]-iV_{0}[\Theta(x)-\Theta(x-a)]\}$, for example. The threshold for $V_0$ is determined when the maximum of the common ratio in Eq.(\ref{t}) equals $1$; namely,
\begin{align}
|r_1 r_2 \exp (2ik'a)|_{E=0}=1 \label{eq:critical}
\end{align}
where
\begin{align}
\notag
&r_{1}=\frac{\sqrt{E}-\sqrt{E-U_{0}-iV_0}}{\sqrt{E}+\sqrt{E-U_{0}-iV_0}}\\
\notag
&r_{2}=\frac{\sqrt{E-U_{0}-iV_0}-\sqrt{E-U_0+iV_0}}{\sqrt{E-U_{0}-iV_0}+\sqrt{E-U_0+iV_0}}\\
&k'=\sqrt{2m(E-U_{0}-iV_{0})}.
\end{align}
For the parameters $m=0.5$, $a=1$ and
$U_{0}=50$, this critical $V_0$ equals $0.00136$.
Numerical calculations show that this is the same threshold for the transmission amplitude to exceed 1, which shall be proved in Section III to be the harbinger for the bifurcation of specturm for the \emph{PT}-symmetric periodic potential. 

\subsection{Evolution of a wave packet through the imaginary barrier}
When a wave packet impinges on a real potential barrier, the
resonances are not critical to its evolution since their
transmission amplitude is of the same order as the nonresonant ones.
However, as is shown in Tab.~\ref{tb:table}, the resonant amplitudes
can reach a few tens of a thousand for an imaginary barrier. It is,
therefore, of interest to study how a wave packet progresses through
such a potential.

As usual, we first decompose it into the scattering states
$\phi_{k}$ in Eq.(\ref{phi}) and the localized states $\phi_{n}$ in
Eq.(\ref{eq:fstate}):
\begin{align}
\Psi(x,t=0)=\sum_{k}C_{k}\phi^{R}_{k}(x,0)+\sum_{n}C_{n}\phi^{R}_{n}(x,0)
\label{expansion}
\end{align}
where the superscript $R$ denotes the right eigenstate. This is
similar to the case of nonHermitian matrices, for which the right
and left eigenstates may not be the same.  Since they can exhibit
different physical properties and time evolution, it is important to
distinguish them. Take the imaginary square potential, for instance.
The barrier seen by the right eigenstates in Eqs.(\ref{phi}) and
(\ref{eq:fstate}) will turn into a well for their left counterparts,
which are defined by
\begin{equation}
\langle\phi^{L}_{k/n}|\hat{H}=\langle\phi^{L}_{k/n}|E_{k/n}
\label{left}
\end{equation}
or
\begin{equation}
 \Big\{\frac{\hat{p}^{2}}{2m}-iV_{0}\big[\Theta(x)-\Theta(x-a)\big]\Big\}\phi^{L}_{k/n}=E^{*}_{k/n}\phi^{L}_{k/n}
\end{equation}
when expressed in the real space. These two eigenstates obey the
following orthogonality relation
\begin{align}
\big\langle\phi^{L}_{k/n}|\phi^{R}_{k'/n'}\big\rangle=\big\langle\phi^{L}_{k/n}|\phi^{R}_{k/n}\big\rangle\delta_{k/n,k'/n'}.
\end{align}
By use of the above relation and the general product, \emph{c}
method, introduced by Gilary {\it et al.}\cite{evolve}, the
weightings $C_{k}$, $C_{n}$ in Eq.(\ref{expansion}) can be
determined as
\begin{align}
C_{k/n}
=\frac{\int \big[ \phi^{L}_{k/n}(x)\big]^*\Psi(x,t=0)dx }{\int\big[
\phi^{L}_{k/n}(x)\big]^* \phi^{R}_{k/n}(x)dx} \label{eq:c-coeff}
\end{align}

\begin{figure}
\includegraphics[width=0.4\textwidth]{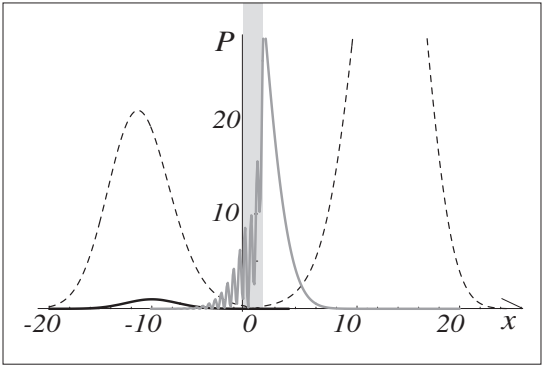}
\caption{The evolution of the wave packet in Eq.(\ref{packet})
impinges on an imaginary square barrier in the gray area with
parameters described in the text. The solid, gray, and dashed lines
represent the distribution at $t=0$, 2, and 5, respectively.}
\label{f:pac}
\end{figure}

The initial wave packet, $\Psi(x,0)$, is chosen as a harmonic
coherent state:
\begin{align}
\Psi(x,0)=\Big({2b\over \pi}\Big)^{1\over 4} e^{
-b(x-x_{0})^{2}+ip_{0}(x-x_{0})} \label{packet}
\end{align}
where $x_{0}$ and $p_0$ denote the position and momentum of the
center of the packet, while $1/2\sqrt{b}$ is its mean width. For an
imaginary square barrier of height $V_{0}=5$, the evolution of wave
packet is illustrated in Fig.\ref{f:pac} with $(x_{0},b)=(-10,0.08)$
and $p_{0}=5.23$ chosen to be near one of the resonance peaks in
Fig.\ref{f:tr0}. It can be seen that the amplitude gets enhanced
upon entering the imaginary barrier due to the resonance. The
oscillatory behavior only exists at intermediate time, as depicted
by the gray line. As the wave packet exits the barrier, the
amplitudes of both transmitted and reflected waves have already been
enhanced by $55$ and $20$ times each.

\section{\emph{PT}-SYMMETRIC PERIODIC POTENTIAL}
\subsection{Origin of bifurcation}
In the following, we generalize the single imaginary potential to a
\emph{PT}-symmetric periodic one. Set the unit cell in $-d/2 \leq x
\leq d/2$ and denote the eigenfunction by
\begin{equation}
\Phi (x)=\phi_{r,k}(x)+A \phi_{l,k}(x) \label{u1}
\end{equation}
where $\phi_{r,k}(x)$ and $\phi_{l,k}(x)$ are the right- and
left-moving scattering states in the unit cell similar to
Eq.(\ref{phi}). According to the Bloch's theorem, the wavefunction
in its previous cell $-3d/2 \leq x \leq -d/2$ can be represented as:
\begin{equation}
\Phi(x)=e^{-iKd} \big[ \phi_{r,k}(x+d)+A \phi_{l,k}(x+d) \big]
\label{u2}
\end{equation}
where $K$ is the Bloch wave number. Matching the boundary conditions
at $x=-d/2$ gives
\begin{align}
\notag
&  \phi_{r,k}(-\frac{d}{2})-e^{-iKd}\phi_{r,k}(\frac{d}{2})=A \big[e^{-iKd}\phi_{l,k}(\frac{d}{2})-\phi_{l,k}(-\frac{d}{2})\big]\\
& \phi'_{r,k}(-\frac{d}{2})-e^{-iKd}\phi'_{r,k}(\frac{d}{2})=A
\big[e^{-iKd}\phi'_{l,k}(\frac{d}{2})-\phi'_{l,k}(-\frac{d}{2})\big]
\label{u3}
\end{align}
where $\phi^\prime$ denotes $d\phi/dx$. By use of the equality
$[\phi_{r,k}\phi'_{l,k}-
\phi'_{r,k}\phi_{l,k}]|^{x=d/2}_{x=-d/2}=0$, the above equations
require the following condition in order to have a nonzero $A$:
\begin{equation}
\frac{U_{1}-U_{2}}{2\big[\phi_{r,k}\phi'_{l,k}-
\phi'_{r,k}\phi_{l,k}\big]|_{x=d/2}}=\cos Kd \label{band-eq}
\end{equation}
where
\begin{equation}
\notag
U_{1}=\phi_{r,k}(-\frac{d}{2})\phi'_{l,k}(\frac{d}{2})+\phi_{r,k}(\frac{d}{2})\phi'_{l,k}(-\frac{d}{2})
\end{equation}
and  the definition of $U_{2}$ is similar but with the subindices
$r$ and $l$ interchanged. Equation (\ref{band-eq}) gives us the
dispersion relation for the spectrum.

Let us write down the form of $\phi_{r,l}$ at $x=\pm d/2$ explicitly
\begin{align}
\notag &\phi_{r,k}(x)= \left\{\begin{array}{ll}
e^{ikx}+R_{r}e^{-ikx}, &x=-\frac{d}{2} \\
T e^{ikx},   &x=\frac{d}{2}
\end{array} \right.\\
&\phi_{l,k}(x)= \left\{\begin{array}{ll}
T e^{-ikx}, &x=-\frac{d}{2} \\
e^{-ikx}+R_{l}e^{ikx},   &x=\frac{d}{2}
\end{array} \right.
\label{ptstates}
\end{align}
where the amplitudes $R,T$ may vary with $k$. Inserting them in Eq.(\ref{band-eq}) gives the relation between the
energy $E=k^2/2m$ and $K$ for a general periodic potential:
\begin{equation}
\frac{T^{2}-R_{r}R_{l}}{2 T}e^{ikd}+\frac{1}{2T}e^{-ikd}=\cos Kd.
\label{dispersion}
\end{equation}

The fact that we are interested at the branch points of
\emph{PT}-symmetric periodic potentials allows us to further
simplify Eq.(\ref{dispersion}). First, take the complex conjugate of
the Schr\"{o}dinger equation for $\phi_r$
\begin{equation}
\Big[\frac{-1}{2m}\frac{d^{2}}{d
x^{2}}+V^{*}(x)\Big]\phi^{*}_{r,k}(x)=E^{*}\phi^{*}_{r,k}(x).
\label{r-phi*}
\end{equation}
If we choose the unit cell to also exhibit the \emph{PT}-symmetry,
the potential $V^{*}(x)$ will become identical to $V(-x)$.
Furthermore, $E^{*}$ is the same as $E$ since the eigenenergy is
real at the branch point. These two properties combined with
Eq.(\ref{r-phi*}) tell us that $\phi^{*}_{r,k}(x)$ must be a linear
combination of the two scattering states:
\begin{align}
\phi^{*}_{r,k}(x)=B \phi_{l,k}(-x)+C \phi_{r,k}(-x). \label{phi*phi}
\end{align}
This relation gives four equations, each of which corresponds to
matching the right- and left-moving part of the wavefunctions on
either side of the unit cell: $R^{*}_{r}=B$, $1=B R_{l}+C T$, $0=B
T+ C R_{r}$ and $T^{*}=C$. It is straightforward to show that they
yield $T^{*}=T/(T^{2}-R_{r}R_{l})$ which simplifies
Eq.(\ref{dispersion}) to:
\begin{equation}
\frac{1}{|T|}\cos(kd+\theta)=\cos Kd
 \label{pt-band-bif}
\end{equation}
where $\theta$ is the phase of $T$.
If $|T|$ should ever be larger than 1 near the extrema of the LHS in Eq.(\ref{pt-band-bif}), there will be no real solution for those $K$ that make $|\cos Kd|$ large.
In other words, the eigenenergies near the Brillouin zone
boundaries of $K$ are sure to be complex. We emphasize that the condition that $|T|$ exceeds unity for some $k$ alone is not enough to predict the bifurcation. It has to coincide with the occurrence of extreme values for $\cos(kd+\theta)/|T|$. This is complement to the previous conclusion\cite{visual} that perturbation forbids bifurcation. What we have shown here is that $|T|\le 1$ guarantees there will be no bifurcation. It is then natural to ask whether $|T|\le 1$ implies the validity of perturbation. As far as the \emph{PT}-symmetric periodic square potential in the next subsection is concerned, this statement has been checked to hold in Section II(B). 

\subsection{\emph{PT}-symmetric periodic square potential}

\begin{figure}[h!]
\includegraphics[width=0.4\textwidth]{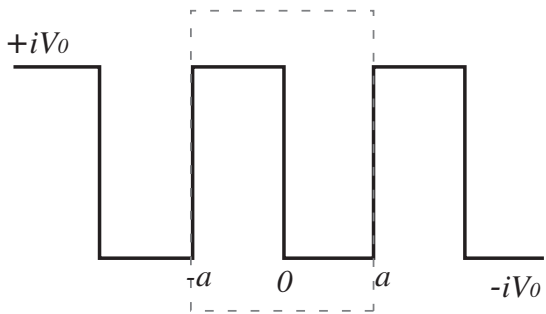}
\caption{A \emph{PT}-symmetric periodic potential with its unit cell
highlighted by the dashed line. Parameters are set at $V_{0}=5$,
$a=1$ and $m=1$.} \label{f:pt}
\end{figure}

As a demonstration, we quantify our conclusions in the previous subsection for the special case of a \emph{PT}-symmetric periodic square
potential, as shown in Fig.\ref{f:pt}. 
The transimission coefficient for the unit cell can be calculated as:
\begin{align}
T=\frac{t_{a}t_{b}t_{c}e^{i(k^{+}+k^{-}-2k)a}}
{1-r_{a}r_{b}e^{2ik^{+}a}
+r_{b}r_{c}e^{2ik^{-}a}-r_{a}r_{c}e^{2i(k^{+}+k^{-})a}} \label{ptt}
\end{align}
where $k=\sqrt{2mE}$ and $k^{\pm}=\sqrt{2 m (E\mp iV_0)}$ represents the momentum in the imaginary barrier/well region. The transmission and reflection amplitudes corresponding to the three boundaries are denoted by $t_{a}=2k/(k+k^{+})$, $t_{b}=2k^{+}/(k^{+}+k^{-})$,
$t_{c}=2k^{-}/(k^{-}+k)$, $r_{a,b}=1-t_{a,b}$ and
$r_{c}=t_{c}-1$.

\begin{figure}
\includegraphics[width=0.45\textwidth]{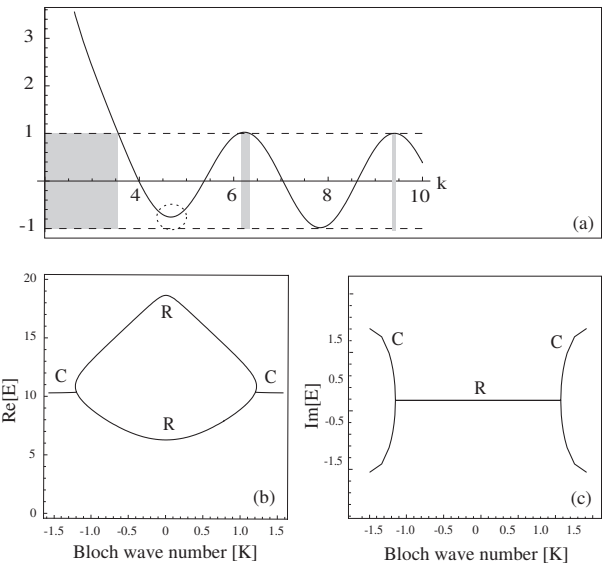}
\caption{(a) The RHS of Eq.(\ref{pt-band-bif}),
$\cos(kd+\theta)/|T|$, is shown in the solid line for
the potential and parameters defined in Fig.\ref{f:pt}. There will be allowed (unshaded)
and forbidden (shaded) regions of $k$. When $\cos Kd$ intersects the point highlighted by
the dashed circle, the corresponding $K$ value illustrates one branch point. (b) The real part of the first
dispersion band. The capital letter R/C indicates that the spectrum
is real/complex in magnitude.
 (c) The imaginary part of the first dispersion band. }
\label{f:band-k}
\end{figure}

Inserting Eq.(\ref{ptt}) into Eq.(\ref{pt-band-bif}) and the
dispersion relation can be determined as:
\begin{align}
\cos (k^{+}+k^{-})a-\frac{(k^{+}-k^{-})^2 }{2k^{+}k^{-}}\sin
k^{+}a \sin k^{-}a=\cos 2Ka. \label{eq:restriction}
\end{align}
The band structure is solved numerically and plotted in
Fig.\ref{f:band-k}.

The perturbation method has been demonstrated to break down for a single localized complex barrier potential in Section II(B), either with or without an accompanying well. When it is generalized to
a \emph{PT}-symmetric periodic potential, we are sure that the
perturbation remains at fault from the analysis within one
unit cell. In the meantime, Fig.\ref{f:band-k} shows that the
originally real eigenenergies evolve into both complex
and real spectra. So the existence of branch points and the
breakdown of perturbation are tied together as proved by
Ref.\cite{visual}. In the mean time, the values of $|T|$ at these extrema also become greater than unity, as is shown in Tab.\ref{tb:lattice}. 

In passing, we observed a special bond between the branch points and the Fano resonances discussed in Section II(B).
Since the momenta $k^+$ and $k^-$ here
happen to be complex conjugates when $E$ is real, the reflection amplitude in
Eq.(\ref{phase}) is rendered pure imaginary. The condition for constructive
interference becomes identical to Eq.(\ref{n}).  This is supported by the numerical results for the branch points presented in
Tab.\ref{tb:lattice}, which indeed approach those for the resonances. 


\begin{table}[tbp]
\begin{tabular}[t]{|c|c|c|c|c|}
\hline
$E$ & 10.39 & 30.62 & 60.34 & 99.87\\
\hline
$|T|$ & 1.36 & 1.02 & 1.00 & 1.00\\
\hline
$n+1/2$ & 1.50 & 2.50 & 3.50 & 4.50\\
\hline
\end{tabular}
\caption{Partial list of energies and absolute magnitude of the transmission amplitude at the
branch points. Their corresponding resonance numbers $n$ from
$n+1/2=2a/\lambda=a \sqrt{2mE}/\pi$ indeed are close to integers.} \label{tb:lattice}
\end{table}

\section{DISCUSSIONS AND CONCLUSIONS}

 Since the non-Hermitian potential results from our examining an open system which is allowed to exchange particles with its environment, the conventional continuity equation is no longer obeyed\cite{pt-sq1,pt-sq3,non-unitarity1,non-unitarity2}. One interesting debate\cite{localize} is whether it is meaningful to mend this loss of unitarity.
Several studies\cite{localize,delta-perturb,zno} have been devoted
to this effort. They tried to construct a metric and corresponding
transformed wave function as the general
framework\cite{n1,n2,np1,np2,np3} to examine the subsystem in a
quasi-Hermitian analysis. Although the unitarity can be restored,
one is constrained to discuss wavefunctions either coming
simultaneously from both sides\cite{localize,delta-perturb} or
exhibiting different normalization constants on either side of the
potential\cite{zno}. To resolve these dilemmas,
Jones\cite{delta-perturb} suggested that one simply treats the
non-Hermitian scattering potential as an effective one in the
standard framework of quantum mechanics. This is the view we
adopt in this work at studying the localized NH
potentials as a phenomenological model of a quantum sink or source
in an open system.

In conclusion, although the breakdown of perturbation approach has
been identified to be intimately linked to the existence of complex
spectrum for a \emph{PT}-symmetric periodic potential,  we offer new
insights into their relationship by studying the less sophisticated
case of an imaginary barrier potential. Due to the simplicity of
this potential form, we are able to elucidate the cause of the
similar breakdown of perturbation, why the perturbation can be
remedied by the superimposition of a sufficiently large real
potential, and the condition for a constructive interference which
differs from that for a real potential. Most importantly, we provide a more comprehensive criteria through Eq.(\ref{pt-band-bif}) on why, when and where the bifurcation will occur.


We thank Hsiu-Hau Lin for useful comments and acknowledge the
support by NSC in Taiwan under grant No. 95-2112-M007-046-MY3.


\begin{thebibliography}{9}

\bibitem{pt1} C. M. Bender and S. Boettcher, Phys. Rev. Lett. \textbf{80}, 5243 (1998).
\bibitem{pt2} C. M. Bender, Contemp. Phys. \textbf{46}, 277 (2005).
\bibitem{16} A. Mostafazadeh, Ann. Phys. (N.Y.) \textbf{309}, 1 (2004).
\bibitem{19} C. M. Bender, Rep. Prog. Phys. \textbf{70}, 947 (2007).
\bibitem{brach1} C. M. Bender, D. C. Brody, H. F. Jones, and B. K. Meister, Phys. Rev. Lett. \textbf{98}, 040403 (2007).
\bibitem{brach2} A. Mostafazadeh, Phys. Rev. Lett. \textbf{99}, 130502 (2007).

\bibitem{dorey} P. Dorey, C. Dunning, and R. Tateo, J. Phys. A: Math. Gen. \textbf{34}, 5679 (2001).
\bibitem{soliton} Z. H. Musslimani, K. G. Makris, R. El-Ganainy, and D. N. Christodoulides, Phys. Rev. Lett. \textbf{100}, 030402 (2008).
\bibitem{beam} K. G. Makris, R. El-Ganainy, and D. N. Christodoulides,and Z. H. Musslimani, Phys. Rev. Lett. \textbf{100}, 103904 (2008).
\bibitem{visual} S. Klaiman, U. G\"{u}nther, and N. Moiseyev, Phys. Rev. Lett. \textbf{101}, 080402 (2008).
\bibitem{pt-sq3} F. Cannata, J-P. Dedonder, and A. Ventura, Ann. Phys. (N.Y.) \textbf{322}, 397 (2007).

\bibitem{evolve} I. Gilary, A. Fleischer, and N. Moiseyev, Phys. Rev. A. \textbf{72}, 012117 (2005).



\bibitem{pt-sq1} Z. Ahmed, Phys. Lett. A. \textbf{324}, 152 (2004).

\bibitem{non-unitarity1} Z. Ahmed, C. M. Bender, and M. V. Berry, J. Phys. A: Math. Gen. \textbf{38}, L627 (2005).
\bibitem{non-unitarity2} M. Znojil, J. Phys. A: Math. Gen. \textbf{39}, 13325 (2006).
\bibitem{localize} H. F. Jones, Phys. Rev. D. \textbf{76}, 125003 (2007).
\bibitem{delta-perturb} H. F. Jones, Phys. Rev. D. \textbf{78}, 065032 (2008).
\bibitem{zno} M. Znojil, Phys. Rev. D. \textbf{78}, 025026 (2008).
\bibitem{n1} C. M. Bender, D. C. Brody, and H. F. Jones, Phys. Rev. Lett. \textbf{89}, 270401 (2002); {\it ibid.} \textbf{92}, 119902(E) (2004).
\bibitem{n2} A. Mostafazadeh, J. Math. Phys. (N.Y.) \textbf{43}, 205 (2002).
\bibitem{np1}H. F. Jones, J. Phys. A: Math. Gen. \textbf{38}, 1741 (2005).
\bibitem{np2}A. Mostafazadeh, J. Phys. A: Math. Gen. \textbf{38}, 6557 (2005);  {\it ibid.} \textbf{38}, 8185(E) (2005).
\bibitem{np3} H. F. Jones and J. Mateo, Phys. Rev. D \textbf{73}, 085002 (2006).
\end{thebibliography}
\end{document}